**НАЦІОНАЛЬНА АКАДЕМІЯ НАУК УКРАЇНИ**
**ІНСТИТУТ МОЛЕКУЛЯРНОЇ БІОЛОГІЇ ТА ГЕНЕТИКИ**

**Дуплій Діана Ростиславівна**

УДК 577.212

# ВЗАЄМОЗВ'ЯЗОК НУКЛЕОТИДНОГО СКЛАДУ ПОСЛІДОВНОСТЕЙ ГЕНОМУ ЛЮДИНИ ЗІ СТРУКТУРОЮ ТА ОСОБЛИВОСТЯМИ ЕКСПРЕСІЇ ГЕНІВ

03.00.03 – молекулярна біологія

**Автореферат**
дисертації на здобуття наукового ступеня
кандидата біологічних наук

КИЇВ – 2010

Дисертацією є рукопис.

Роботу виконано у відділі біохімічної генетики Інституту молекулярної біології та генетики НАН України (м. Київ).

**Науковий керівник:** кандидат біологічних наук, старший науковий співробітник
**Чащин Микола Олексійович**,
Інститут молекулярної біології та генетики НАН України,
старший науковий співробітник відділу біохімічної генетики.

**Офіційні опоненти:** доктор біологічних наук, професор,
член-кореспондент НАН України
**Говорун Дмитро Миколайович**,
Інститут молекулярної біології та генетики НАН України,
заступник директора з наукової роботи,
завідувач відділу молекулярної та квантової біофізики;

кандидат біологічних наук,
старший науковий співробітник
**Карпов Павло Андрійович**,
Інститут харчової біотехнології та геноміки НАН України,
м. Київ,
завідувач лабораторії біоінформатики та структурної біології.

Захист відбудеться 25 січня 2011 року о 10.00 годині на засіданні спеціалізованої вченої ради Д 26.237.01 Інституту молекулярної біології та генетики НАН України за адресою: 03680, Київ, вул. Академіка Заболотного, 150.

З дисертацією можна ознайомитися у бібліотеці Інституту молекулярної біології та генетики НАН України.

Автореферат розіслано ____24__ грудня 2010 року.

Вчений секретар
спеціалізованої вченої ради,
кандидат біологічних наук,
ст. науковий співробітник                                                                       І. В. Крупська

# ЗАГАЛЬНА ХАРАКТЕРИСТИКА РОБОТИ

**Актуальність теми.** Нуклеотидний склад первинної послідовності геному людини є фундаментальною характеристикою і відображає результат еволюційного впливу протягом тривалого часу. Дослідження нуклеотидних співвідношень послідовностей генома є засобом встановлення зв'язку між їхньою первинною структурою і функцією. Залучення обчислювальних методів до аналізу генома дозволяє досліджувати нові зв'язки між структурою генів і генетичними процесами.

За теорією нейтральної еволюції (Kimura, 1985) первинна структура ДНК піддається двом протилежно спрямованим впливам. З одного боку, відбувається дестабілізуюча дія мутаційного тиску, що змінює сенс генетичного тексту, а з іншого – має місце стабілізуючий вплив природного відбору і систем репарації, що підтримують консервативність генетичного коду. Показано, що частоти спонтанного мутагенезу відрізняються не лише серед біологічних видів (Lynch, 2008) – навіть у межах одного геному частота мутацій різних локусів може відрізнятися у тисячу разів (Колотова та Стегній, 2004). Це обумовлено функціональними вимогами до локусу: так, наприклад, екзони перебувають під більш строгим контролем природного відбору в порівнянні з інтронами.

Причинами нуклеотидних замін, крім шкідливої дії відомих мутагенів, є також помилки процесів матричного синтезу, зокрема реплікації і транскрипції (Mugal et. al, 2010). Виявлено, що системи репарації віддають перевагу областям з активною транскрипцією у порівнянні з міжгенними проміжками (Reid and Svejstrup, 2004). Точкові мутації накопичуються протягом філогенезу – деякі з них не мають фенотипічних проявів, а лише забезпечують генетичне різноманіття. Наслідком таких мутацій є сайти однонуклеотидних поліморфізмів (Messer, 2009). У той же час будь-які генеративні мутації на рівні організму піддаються дії відбору. Оскільки геном людини пройшов тривалий еволюційний шлях, то його первинна структура містить ознаки і наслідки еволюційного впливу.

Правило про рівну кількість комплементарних нуклеотидів у нативній ДНК встановлено експериментально (Chargaff, 1952), але його сенс став зрозумілим лише після встановлення просторової структури ДНК (Watson and Crick, 1953). Пізніше було виявлено, що це правило поширюється на кожний ланцюжок ДНК (Lin and Chargaff, 1967). Таким чином, сформульовано два правила комплементарності нуклеотидів – для дво- і одноланцюгової ДНК відповідно (Sueoka, 1995). Правило комплементарності для одноланцюгової ДНК не випливає із правила Чаргаффа, яке не накладає ніяких обмежень на частоти зустрічальності чотирьох нуклеотидів (A, T, G, C) в одному ланцюзі ДНК. Причини цього феномену до сьогоднішнього дня так і не з'ясовано. Вже перші дослідження секвенованих ДНК продемонстрували статистичну нерівномірність розподілу моно- і динуклеотидів (Nussinov, 1984; Bulmer, 1987). Пізніше масштабні біоінформатичні обчислення довели, що частоти комплементарних нуклеотидів у межах геномів і хромосом вздовж одного ланцюжка ДНК майже однакові у багатьох видів вищих евкаріотів (Сергієнко та Гупал, 2005). У той же час на менших ділянках нуклеотидних послідовностей кількості комплементарних нуклеотидів не збігаються або ж спостерігається асиметрія нуклеотидного складу (Francino and Ochman, 1997; Mugal et al., 2009). Виникає запитання – чим обумовлені нуклеотидні асиметрії одноланцюгової ДНК?

За даними літератури біоінформатичні підходи щодо дослідження асиметрії нуклеотидного складу ґрунтуються на двох різних ідеях. Перший із них спрямований на філогенетичну реконструкцію нуклеотидних замін шляхом вирівнювання і аналізу гомологічних послідовностей від одного спільного попередника (O'Meara, 2008). Такі дослідження ускладнюються пошуком послідовностей попередників, оскільки шляхи дивергенції ортологічних генів складні та заплутані (Kuzniar et al., 2008). Інший підхід ґрунтується на обчисленні показників асиметрії нуклеотидного складу (skew), що, зазвичай розраховуються як відношення різниці частот зустрічальності комплементарних нуклеотидів до їхньої суми (Frank and Lobry 1999; Touchon et al., 2005). Попри свою відносну простоту, цей підхід виявився доволі інформативним. Саме завдяки йому спрогнозовано нові точки реплікації (Ma et al. 2004; Touchon et al., 2005), а також встановлено зв'язок асиметрій із процесом транскрипції (Polak et. al., 2010; Fujimori et al., 2005) тощо.

Дисертаційне дослідження спрямоване на виявлення залежності складу послідовності гена від його функції та еволюційного впливу. Виявлення та систематизація таких залежностей дозволить наблизитися до розуміння механізмів спадковості та мінливості.

**Зв'язок роботи з науковими програмами, планами, темами.** Дисертаційну роботу виконано у рамках наукових проектів відділу біохімічної генетики Інституту молекулярної біології та генетики НАН України «Пошук структурно-функціональних взаємозв'язків цитохрому Р450» (НДР НАН України, № держ. реєстрації – 0198U000675, 1998-2003 рр.) і «Створення нової моделі тестування гепатотоксичності шкідливих чинників навколишнього середовища людини» (НДР НАН України «Новітні медико-біологічні проблеми та навколишнє середовище людини», № держ. реєстрації – 0106U000537, 2004-2006 рр.).

**Мета та завдання дослідження.** Мета дослідження – виявити взаємозв'язок нуклеотидного складу послідовностей геному зі структурою генів та особливостями їхньої експресії у людини, а також проаналізувати відмінності у послідовностях генів цитохрому Р450 2Е1 людини та деяких ссавців.

Для досягнення цієї мети необхідно було вирішити такі завдання:
1. Розробити програмні реалізації щодо оптимізації роботи з нуклеотидними послідовностями.
2. Дослідити чисельні показники нуклеотидного складу для екзонів, інтронів та ділянок, що оточують ген. Встановити взаємозв'язок асиметрій нуклеотидного складу з інтронно-екзонною структурою генів.
3. Виявити взаємозв'язок асиметрій нуклеотидного складу з характером та рівнем експресії генів за літературними даними аналізу ДНК-чипів.
4. Проаналізувати чисельні показники нуклеотидного складу для генів, що експресуються конститутивно і виконують базові клітинні функції, та для тканиноспецифічних генів.

5. Провести порівняльний аналіз нуклеотидних послідовностей генів цитохрому Р450 2Е1 людини та деяких ссавців.

*Об'єкт дослідження* – композиційні закономірності нуклеотидного складу геному людини, пов'язані з різними структурними елементами генів.

*Предмет дослідження* – чисельні показники зустрічальності нуклеотидів та нуклеотидні співвідношення у геномі людини.

*Методи дослідження* – біоінформатичний аналіз нуклеотидних послідовностей, структурне програмування, статистичний аналіз результатів, візуалізація числових масивів даних.

**Наукова новизна отриманих результатів.** Вперше проведено порівняльний аналіз величин асиметрій генів людини і прилеглих до них областей. Вперше встановлено взаємозв'язок між величинами асиметрій нуклеотидних послідовностей генів та рівнем їхньої експресії у людини. Розроблено програмний інструментарій для аналізу послідовностей бази NCBI і створено власну базу даних мРНК екзонів, інтронів, а також областей, прилеглих до 3'- і 5'- кінців генів людини. Зафіксовано суттєву відмінність асиметрій нуклеотидного складу екзонів і інтронів генів людини. В екзонах виявлено вищі показники пуриново-піримідинової, а в інтронах – кето-амінної асиметрії. Зафіксовано залежність величин асиметрій від характеру експресії генів: найвищі значення асиметрій нуклеотидного складу спостерігаються у генах, що експресуються конститутивно і виконують базові клітинні функції. Отримано позитивну кореляцію показника кето-амінної асиметрії з рівнем експресії генів у лінії статевих клітин. Проведено порівняльний аналіз послідовностей генів цитохрому Р450 2Е1 людини та різних ссавців. Зафіксовано накопичення протягом еволюції транзицій CG → TA в інтронах генів *Cyp2e1*.

**Практичне значення результатів.** Результати розрахунків асиметрій нуклеотидного складу можуть бути використані для передбачення функцій невідомих послідовностей. Створено оригінальну базу послідовностей мРНК генів людини, доступну для користувачів. Розроблений дисертантом програмний інструментарій для аналізу послідовностей ДНК геному людини у форматі бази даних NCBI можна застосовувати для дослідження геномів інших біологічних видів.

Матеріали дисертації можуть бути використані у спецкурсах з молекулярної біології та структурної біоінформатики для студентів біологічних факультетів вищих навчальних закладів.

**Особистий внесок здобувача:** пошук і аналіз літературних джерел, розрахунки, одержання, обробка та теоретичне обґрунтування результатів досліджень виконано дисертанткою особисто або за безпосередньої її участі. Здобувачем особисто обрано критерії вибору фрагментів послідовностей ДНК та проведено розрахунки нуклеотидних частот і співвідношень. Програмний інструментарій мовою PERL створено спільно з В. В. Калашніковим (Харківський національний економічний університет). Автор брала безпосередню участь у підготовці символьних послідовностей та налагодженні оригінальних програм.

**Апробація результатів дисертації.** Положення роботи доповідалися і були представлені на вітчизняних і міжнародних конференціях та з'їздах: IV з'їзді Українського біофізичного товариства (Донецьк, Україна 2006), 15th Annual International Conference on Intelligent Systems for Molecular Biology (ISMB) and 6th European Conference on Computational Biology (ECCB) (Vienna, Austria 2007), 4-ій Міжнародній конференції з $p$-адичної математичної фізики (Гродно, Білорусь 2009).

**Публікації.** Основні результати дисертаційної роботи опубліковано у 7 наукових працях, із них – у 2 статтях у фахових журналах, 4 тезах доповідей національних і міжнародних наукових конференціях і з'їздах, захищено 1 патентом України.

**Структура та обсяг дисертації.** Дисертація складається із вступу, огляду літератури, матеріалів і методів дослідження, експериментальної частини, яка складається з трьох розділів, узагальнення результатів дослідження, висновків та переліку використаних джерел, що налічує 154 найменувань. Текст дисертації викладено на 124 сторінках машинописного тесту, вона містить 10 таблиць та 39 рисунків.

## ОСНОВНИЙ ЗМІСТ РОБОТИ

**Матеріали і методи дослідження.** Нуклеотидні послідовності та списки генів отримано з електронних баз даних NCBI (RefSeq, Gene, Nucleotide, Protein, Unigene і GEO). Базу GenAtlas використовували як джерело додаткової і незалежної інформації щодо синонімічних імен генів та їхнього картування тощо. Для створення власної бази даних застосовано методи структурного програмування мовою PERL (Wall et al. 2004). Для розрахунків нуклеотидних замін використовували мову програмування PHP (Котеров и Костарев, 2006). Біоінформатичний пошук подібностей послідовностей проводили за допомогою програмних пакетів BLASTN і BLASTP. Попарне вирівнювання послідовностей виконували за допомогою програми ClustalW. Для множинного вирівнювання та побудови кладистичних дерев застосовували пакет MEGA4 на локальному комп'ютері. Статистичний аналіз результатів проводили за допомогою пакета програм SPSS13.0.

У роботі використано дані геномних послідовностей людини з архіву NCBI. Використовували модель жіночого геному, яка складається з файлів з інформацією про 22 аутосоми та X хромосоми. Для аналізу нуклеотидного складу екзон-інтронних послідовностей генів застосовували формат файлів "gbk".

Формат файлів "fasta" використовували для розрахунків у межах хромосом. Послідовності мРНК вилучали з файлів "gbk", за координатами їхніх функціональних фрагментів. Розбивку файлів контигів ("gbk") здійснювали за допомогою регулярних виразів мови PERL на підставі розширених нормальних форм граматик Бекуса-Науера (Backus, 1959; Aho et al., 2008). Дескриптори генів брали з описів формату файлів банка генетичної інформації (NCBI handbook).

Для дослідження нуклеотидних співвідношень обрано показники асиметрій нуклеотидного складу. Асиметрія у даному випадку означає порушення правила Чаргаффа для одного ланцюжка, тобто відхилення від одиниці співвідношення кількостей комплементарних нуклеотидів. Показники асиметрій нуклеотидного складу розрахували як відношення різниці кількостей комплементарних нуклеотидів до їхньої суми. Чотири типи асиметрій нуклеотидного складу розраховували за формулами:

аденін-тимінову – $AT_{асим} = (NA-NT) / (NA + NT)$;
гуанін-цитозинову – $GC_{асим} = (NG-NC) / (NG + NC)$;
пурин-піримідинову – $R_{асим} = (NA + NG-NT-NC) / L$;
кето-амінну – $K_{асим} = (NG + NT-NA-NC) / L$;

де NA, NG, NT і NC – абсолютні значення кількостей відповідних нуклеотидів на фрагменті довжиною L ($L = NA + NG + NT + NC$).

Для порівняння і оцінки величин асиметрій розраховували середні, зважені середні арифметичні та середньоквадратичні похибки. Достовірність відмінностей величин оцінювали за допомогою критеріїв Фішера. Ряди розподілу величин порівнювали за допомогою критерія Пірсона.

Групи генів з різним характером експресії формували згідно з опублікованим даними про гени, що еспресуються конститутивно і виконують базові клітинні функції (Eisenberg and Levanon 2003), та тканиноспецифічні гени (Hsiao et al., 2001). Для даних щодо величин експресії генів у лінії статевих клітин (сперматоцитів) використовували літературні дані з аналізу ДНК-чипів (GEO ID GSM18985 і GSM18986) [http://www.ncbi.nlm.nih.gov/geo/].

Враховуючи, що існують гени, які можуть перекриватися (Gerstein et al., 2007), включаючи в свою структуру одну і ту ж ділянку ДНК, у роботі досліджено взаємозв'язок такої особливості структури з нуклеотидними асиметріями. Порівнювали групу генів, що повністю перекриваються, тобто коли один ген повністю входить до складу іншого, з групою генів, що не перекриваються, себто розташованих у незалежних локусах ДНК. Для аналізу нуклеотидних замін у послідовностях генів цитохрому Р450 2Е1 використовували дані геномних послідовностей генів цитохрому Р450 2Е1 (*Cyp2e1*) людини і семи ссавців: *Pan troglodytes* XM_508139.2, *Bos taurus* NM_174530.2, *Canis lupus* NM_001003339.1, *Equus caballus* NM_001111303.1, *Mus musculus* NM_021282.2, *Rattus norvegicus* NM_031543.1 і *Sus scrofa* NM_214421. Пошук подібності генів проводили з використанням пакету програм BLATN на основі гомології їхніх послідовностей (Altschul et al., 1990). Попарне вирівнювання проводили програмою ClustalW. Кладістичне дерево будували методом зв'язування сусідів NJ (Saitou and Nei, 1987). Сума довжин гілок оптимального дерева дорівнює 0,73912457. Еволюційнні відстані розраховували за допомогою методу максимальної правдоподібності (Tamura et al., 2004) та виражали у кількості підстановок на одну пару основ. Після проведення множинного вирівнювання позиції послідовностей, що містили розриви та пропуски даних, видаляли перед подальшим проведенням кластеризації.

**Результати дослідження та обговорення.**

*Розробка програмного інструментарію для роботи з масивами генів.* За допомогою існуючого інструментарію NCBI (Entrez Gene, MapViewer та ін.) дуже проблематично реалізувати мету дослідження. Отже автором було взято дані геному людини з файлового серверу NCBI і розроблено оригінальний програмний інструментарій, адаптований до вирішення поставлених завдань дослідження на локальному комп'ютері. Розроблений спосіб полягає в тому, що користувач вибирає визначення функціонального фрагмента, послідовність якого (яких) виділятимуть. Потім генерують набір шаблонів з регулярних виразів і ієрархічно застосовують його до даних вхідного файлу NCBI «GBK». Знайдені фрагменти перетворяться відповідно до заданих вимог і зберігаються у вихідному потоці результату. Завдяки цьому підходу створено базу даних генів людини, яка містить інформацію щодо послідовностей екзонів, інтронів та областей, прилеглих до 5'- та 3'- кінців генів (рис. 1). Отримані послідовності мРНК знаходяться у стандартному текстовому форматі, що робить їх доступними для подальшого аналізу за допомогою зовнішніх програм BLAST і ClustalW. Послідовності мають марковані фрагменти екзонів, інтронів, а також прилеглих до генів областей, що дозволяє аналізувати їх окремо.

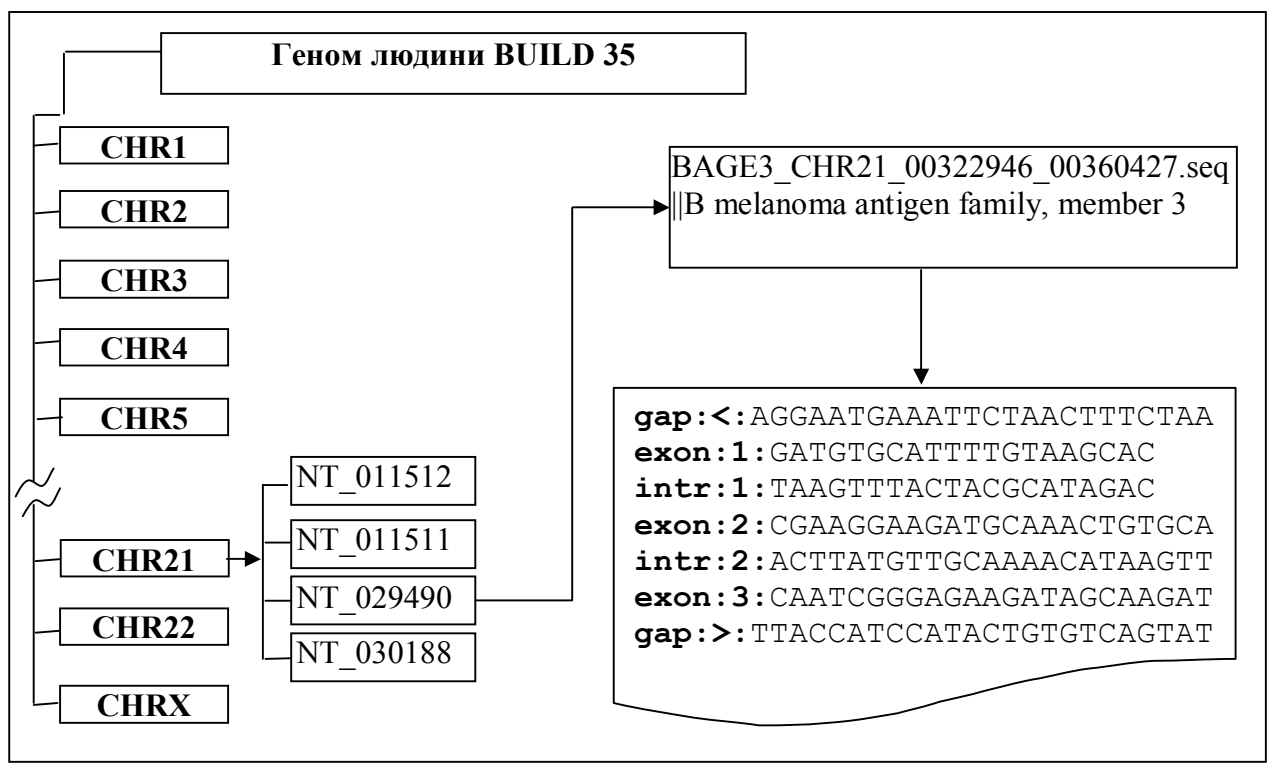

**Рис. 1. Структура бази послідовностей мРНК, створеної дисертантом**

***Взаємозв'язок нуклеотидного складу зі структурою і функцією генів.*** Послідовність кожної хромосоми представлено як набір контигів еухроматичної частини геному. У результаті дослідження повної версії генома людини визначено загальну кількість генів у кожній хромосомі (табл. 1).

*Таблиця 1*

**Кількість контигів, генів і мРНК у геномі людини**

| № хромосоми | Кількість контигів у хромосомі | В тому числі контигів без генів | Кількість генів у хромосомі | У тому числі псевдогенів | Фактична кількість мРНК |
|---|---|---|---|---|---|
| 1 | 47 | 3 | 2422 | 153 | 2269 |
| 2 | 21 | 1 | 1721 | 206 | 1515 |
| 3 | 11 | 3 | 1337 | 118 | 1219 |
| 4 | 21 | 3 | 1000 | 93 | 907 |
| 5 | 7 | 0 | 1163 | 93 | 1070 |
| 6 | 11 | 0 | 1357 | 157 | 1200 |
| 7 | 16 | 2 | 1337 | 196 | 1141 |
| 8 | 18 | 8 | 906 | 70 | 836 |
| 9 | 50 | 12 | 1058 | 115 | 943 |
| 10 | 20 | 3 | 962 | 78 | 884 |
| 11 | 8 | 0 | 1629 | 273 | 1356 |
| 12 | 9 | 0 | 1249 | 98 | 1151 |
| 13 | 6 | 0 | 476 | 59 | 417 |
| 14 | 1 | 0 | 1046 | 368 | 678 |
| 15 | 16 | 2 | 897 | 111 | 786 |
| 16 | 6 | 0 | 977 | 51 | 926 |
| 17 | 16 | 3 | 1320 | 66 | 1254 |
| 18 | 6 | 1 | 395 | 36 | 359 |
| 19 | 6 | 1 | 1570 | 67 | 1503 |
| 20 | 6 | 0 | 668 | 72 | 596 |
| 21 | 5 | 0 | 321 | 49 | 272 |
| 22 | 12 | 0 | 626 | 86 | 540 |
| X | 26 | 4 | 1101 | 146 | 955 |
| Всього | 345 | 46 | 25538 | 2761 | 22777 |

Аналіз довжин функціональних фрагментів показав, що переважна більшість перших екзонів – 90% має довжину до 900 пар нукледотидів (п.н.), причому більше, ніж половина з них (55%) мають довжину від 50 до 250 п.н. (Мода варіаційного ряду = 117,0). Середня довжина першого екзона (373 ± 3,9), більше ніж удвічі перевищує довжину внутрішніх екзонів (168 ± 0,5). Внутрішні екзони у 80% випадків, як і перші, мають довжину від 50 до 200 п.н. Короткі екзони - до 50 п.н. – зустрічаються у 6% серед перших екзонів, у 2,6% з-поміж внутрішніх і лише у 0,3% випадків серед останніх екзонів. Аналіз довжин екзонів 18566 генів засвідчив, що середні довжини екзонів можна розташувати у такому порядку зростання: внутрішні – 169 п.н., перші – 373 п.н. і останні – 1120 п.н. Останні екзони майже втричі довші, ніж перші, і у п'ять разів довші, ніж внутрішні екзони. Серед інтронів пік розподілу припадає на довжину 600 п.н.

Для кожного гена описано до 23 транскриптних варіантів мРНК. Число екзонів в генах значно коливається, хоча середнє число екзонів в одному гені дорівнює приблизно 10, медіана вибірки дорівнює 7 (рис. 2).

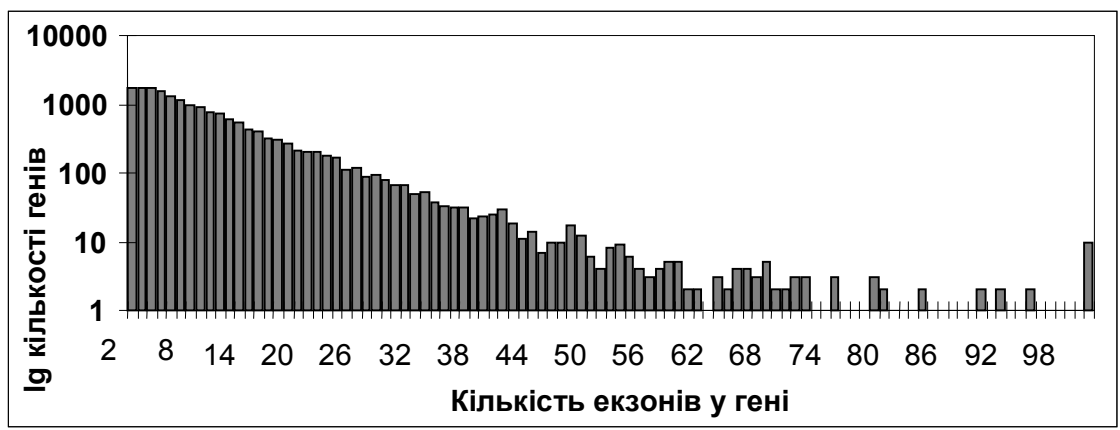

**Рис. 2. Гістограма розподілу кількостей екзонів у 18566 генів людини**

Для поліпшення якості отриманого масиву генів людини проведено сортування і відбір послідовностей для подальшого їхнього аналізу (табл. 2).

*Таблиця 2*

**Етапи сортування масивів генів**

| Дія з сортування | Кількість генів |
|---|---|
| Первинно отримано назв генів з файлів контигів | 25538 |
| Видалено псевдогенів | 2671 |
| Отримано послідовностей мРНК | 22777 |
| Видалено короткі та високогомологічні гени | 19734 |
| Видалено одноекзонні (1169) гени | 18566 |
| Видалено по 10 п.н. на екзонно-інтронних межах | 17830 |
| Отримано результати аналізу позицій кодонів (CDS) | 10839 |
| Отримано дані щодо величин експресії (GEO) | 6472 |

Згідно із сучасним уявленням про структуру евкаріотного гена застосували його загальноприйняту екзонно-інтронну модель побудови. Проаналізовано також суміжні області генів: 1000 п.н. ліворуч від старту транскрипції (5'-кінця), а також 1000 п.н. праворуч від 3'-кінця транскрипту. Порівняльний аналіз середніх величин асиметрій нуклеотидного складу цих фрагментів виявив наявність стійких та істотних відмінностей (рис. 3).

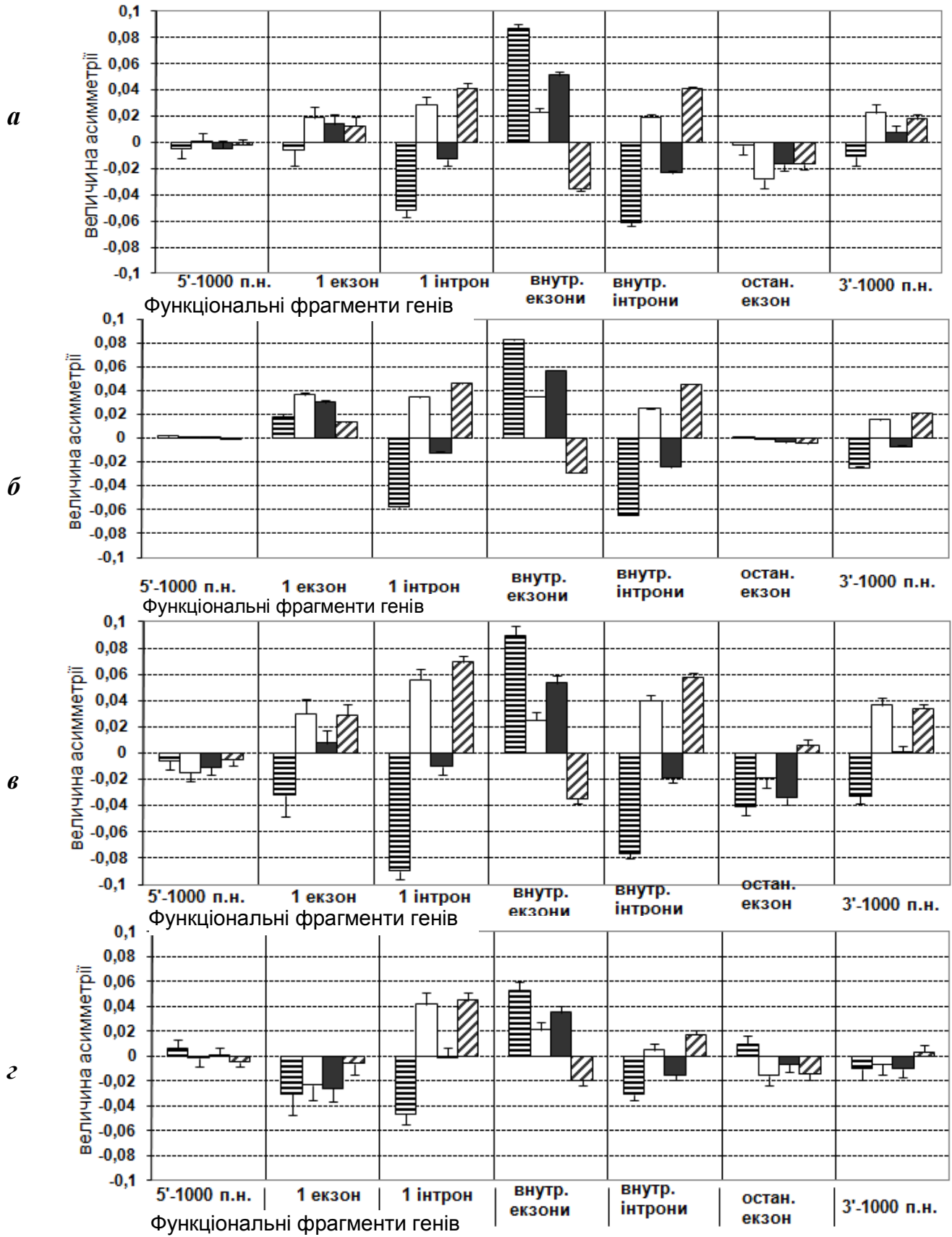

**Рис. 3.** Середні величини асиметрій нуклеотидного складу у різних групах генів: *а* - гени, що не перекриваються; *б* - гени, що повністю перекриваються; *в* – конститутивні; *г* - тканиноспецифічні гени. Горізонтальним штрихуванням позначено величини $AT_{асим}$, білим кольором – $GC_{асим}$, чорним – $R_{асим}$, діагональним штрихуванням – $K_{асим}$

У результаті проведених досліджень отримано величини асиметрій нуклеотидного складу екзонів, інтронів і прилеглих до генів областей. Суміжні області генів: не більше, ніж 1000 п.н. за умови відсутності іншого гена ліворуч від старту транскрипції (5'-кінця), та не більше, ніж 1000 п.н. праворуч від 3'-кінця транскрипту за тих же умов.

Перш за все можна зазначити, що області, які не транскрибуються, мають найменші величини асиметрій. Примітним є і те, що величини асиметрій нуклеотидного складу внутрішніх екзонів і інтронів мають протилежні знаки. Внутрішні інтрони характеризуються негативними показниками $AT_{асим}$ (-0,0446±0,0007) та $R_{асим}$ (-0,0153±0,0005), а екзони – позитивними $AT_{асим}$ (0,0705±0,0012) та $R_{асим}$ (0,0397±0,0009). Протилежні відмінності спостерігаються для $K_{асим}$: в екзонах він дорівнює -0,0321±0,0007, а в інтронах – складає 0,0331±0,0004. Показник $GC_{асим}$ позитивний для всіх генних фрагментів, але в інтронах він (0,0198±0,0006) дещо вищий, ніж в екзонах (0,0145±0,0010). Таким чином, можна зробити висновок про те, що найбільші величини асиметрії нуклеотидного складу спостерігаються в інтронах. Відомо, що природній відбір переважно спрямований на оптимізацію трансляції (Sueoka, 1992). Завдяки виродженості генетичного коду майже усім амінокислотам відповідає кілька кодонів, що відрізняються у більшості випадків за третьою позицією. Отже, точкова мутація у третій позиції кодона, за умови, що вона не порушуватиме його зміст, може залишитися і не буде елімінована відбором. Таким чином, третя позиція кодонів, як і інтрони, можуть накопичувати місенс-мутації. На рис. 4 показано, що для третьої позиції кодонів у різних групах генів людини величина $K_{асим}$ в інтронах вища, аніж $K_{асим}$.

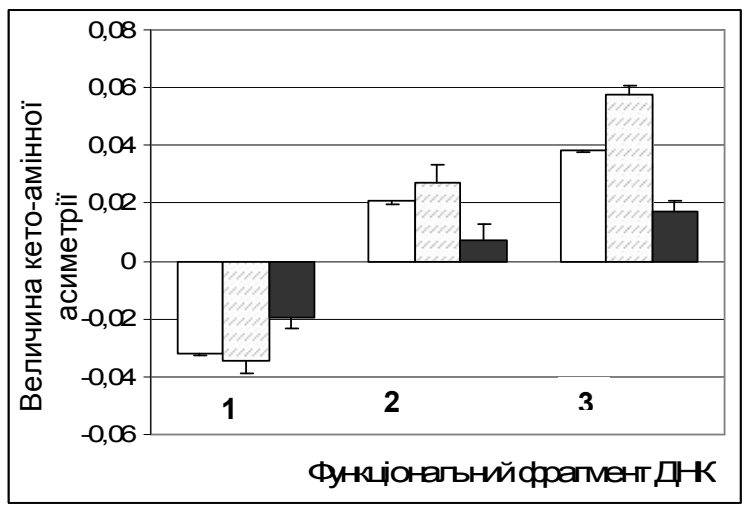

**Рис. 4. Показник кето-амінної асиметрії для внутрішніх екзонів (1), інтронів (2) та третьої позиції кодонів (3). Білий колір - загальна група генів, чорний – група тканиноспецифічних генів, штрихування – група генів, що експресуються конститутивно і виконують базові клітинні функції**

Слід зазначити, що розподіл величин асиметрій вздовж функціональних фрагментів генів статистично однорідний (рис. 5).

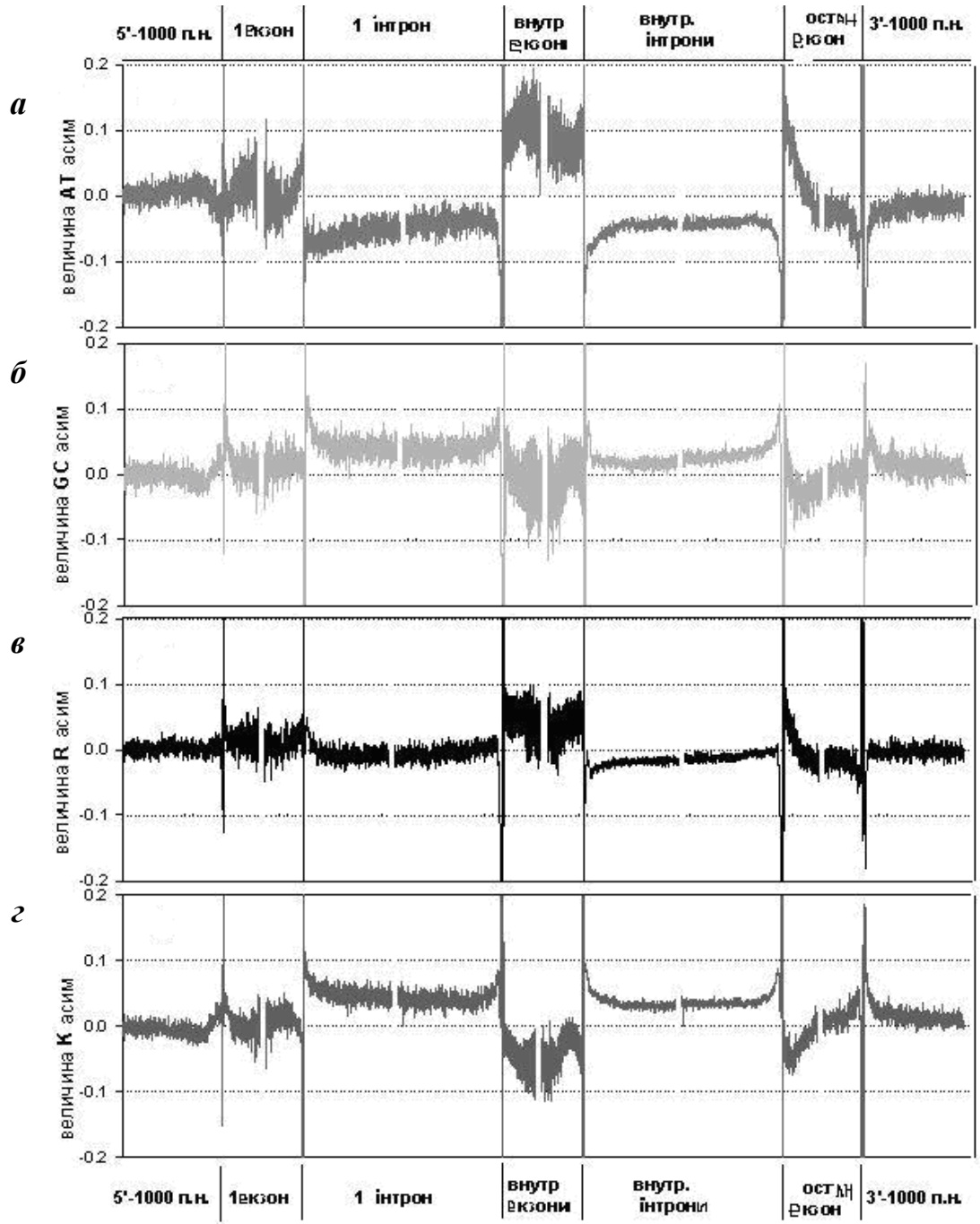

**Функціональні фрагменти ДНК**

**Рис. 5. Розподіл величини асиметрії вздовж гена (крок – 1 нуклеотид):** *а* - аміно-тимінова (AT$_{асим}$); *б* - гуанін-цитозінова (GC$_{асим}$); *в* - пурин-піримідинова (R$_{асим}$); *г* - кето-амінна (K$_{асим}$)

На кожному структурному фрагменті гена спостерігаються відхилення асиметрій від середнього значення лише при безпосередньому наближенні до інтрон-екзонних меж. Можливо, це є наслідком присутності на останніх регуляторних елементів, наприклад сайтів сплайсингу AG, які ми не врахували при отриманні середніх величин асиметрій. Як видно з рис. 5, переважний склад фрагментів формує середні величини останніх.

Підсумовуючи отримані дані, відмінності нуклеотидного складу інтронів і екзонів можна представити у вигляді системи нерівностей (табл. 3).

*Таблиця 3*

**Співвідношення частот зустрічальності
нуклеотидів в екзонах та інтронах генів людини**

| Фрагменти генів | |
|---|---|
| Екзони | Інтрони |
| $N_A > N_T$; | $N_A < N_T$; |
| $N_G \geq N_C$; | $N_C \leq N_G$; |
| $N_A+N_G > N_C+N_T$; | $N_A+N_G \leq N_C+N_T$; |
| $N_A+N_C \geq N_G+N_T$. | $N_A+N_C < N_G+N_T$. |

Отримані величини асиметрії нуклеотидного складу (значень $AT_{асим}$ і $К_{асим}$ в екзонах) віддзеркалюють факт накопичення нуклеотидних замін AT→ GC і CT→TA в локусах, які знаходяться під меншим тиском природного відбору.

У роботі проаналізовано величини експресії за даними бази GEO різних груп генів у лінії статевих клітин. Виявлено, що середній рівень експресії генів, що експресуються конститутивно, більше ніж втричі вищий (1147±81), ніж у тканиноспецифічних генів (314±31). Рівень експресії генів, що повністю перекриваються (384±44) та не перекриваються (450±10) майже однаковий. У результаті такого аналізу отримано позитивну кореляцію між рівнем експресії досліджуваних генів і величинами кето-амінної асиметрії, яка описується рівнянням y = 0,0027x + 0,0188 (рис. 6).

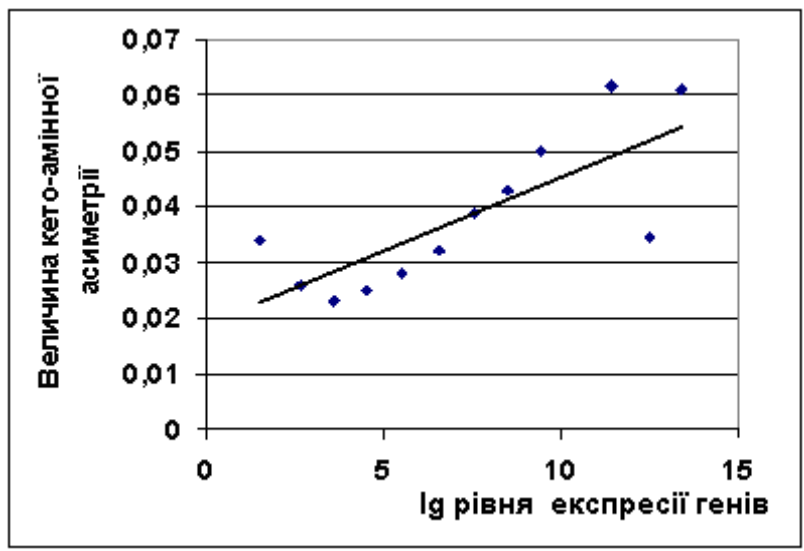

**Рис. 6. Кореляція між величиною кето-амінної асиметрії та рівнем експресії генів у клітинах сперматоцитів людини**

Отримані результати дозволяють зрозуміти можливі механізми формування асиметрій нуклеотидного складу. Відомо, що навіть при незмінності умов зовнішнього середовища причинами спонтанних точкових мутацій можуть бути процеси матричного синтезу, зокрема транскрипція. У той же час за даними літератури відомо, що механізми репарації діють швидше та ефективніше в активних локусах гена і набагато повільніше – в інтронах та міжгенних областях. Гени, які експресуються конститутивно та виконують базові клітинні функції, транскрибуються і експресуються у лінії статевих клітин. Отже, якщо помилки транскрипції у генах цих клітин не елімінуються відбором, то вони успадковуються нащадками. Отримана кореляція між кето-амінною асиметрією інтронів та рівнем експресії гена свідчить про нуклеотидні заміни, які збереглися протягом еволюції. Переважання тиміну в інтронах є, можливо, результатом накопичення у них транзицій CG→TA.

***Аналіз нуклеотидних замін у генах ортологічних білків цитохрому P450 2E1 (CYP2E1)***. Цитохром P450 є унікальним білком, який зазнав еволюційний шлях від примітивних аеробних прокаріотів до людини. За даними літератури попередник цитохрому існував приблизно 2 млрд. років тому (Nelson,1987). У зв'язку з цим у роботі проведено аналіз геномних послідовностей генів однієї з ортологічної ізоформи білка цитохрому P450 2E1 людини і семи ссавців – *Homo sapiens, Bos taurus, Canis lupus, Equs cabbalus, Mus musculus, Rattus norvegicus і Sus scrofa*. Для встановлення подібності первинної структури генів *Cyp2e1* проведено множинні вирівнювання його трансльованих послідовностей. На підставі отриманих даних побудовано кладистичне дерево (рис. 7). Еволюційнні відстані розраховані методом максимальної правдоподібності та представлені як кількость підстановок на одну пару основ.

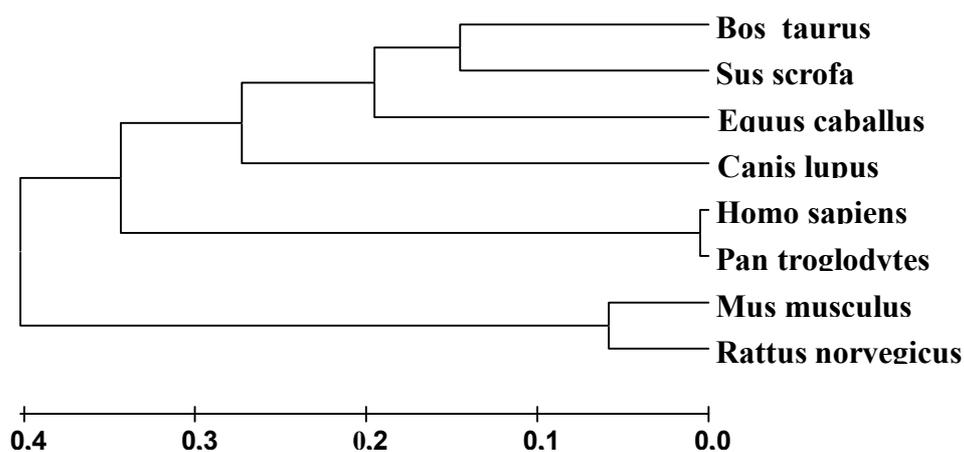

**Рис. 7. Кладистичне дерево послідовностей генів цитохрому P450 2E1**

Як видно з рис. 7, послідовність гена цитохрому P450 2E1 зазнала декількох дивергенцій, зокрема у представників гризунів, приматів та інших ссавців. Виявлено близькість послідовностей *Cyp2e1 Homo sapiens* із послідовностями *Mus musculus* і *Rattus norvegicus*. Це свідчить, що проаналізовані гени *Cyp2e1* мали спільну предкову послідовність. Для того, щоб відслідкувати зміни нуклеотидного складу генів *Cyp2e1* серед ссавців розраховано К$_{асим}$ інтронів і екзонів *Cyp2e1* кожного біологічного виду. Отримані результати порівнювали з раніше отриманими даними для загальної вибірки генів людини. На рис. 8 показано динаміку зміни величини К$_{асим}$ у залежності від еволюційної спорідненості нуклеотидних послідовностей генів *Cyp2e1*.

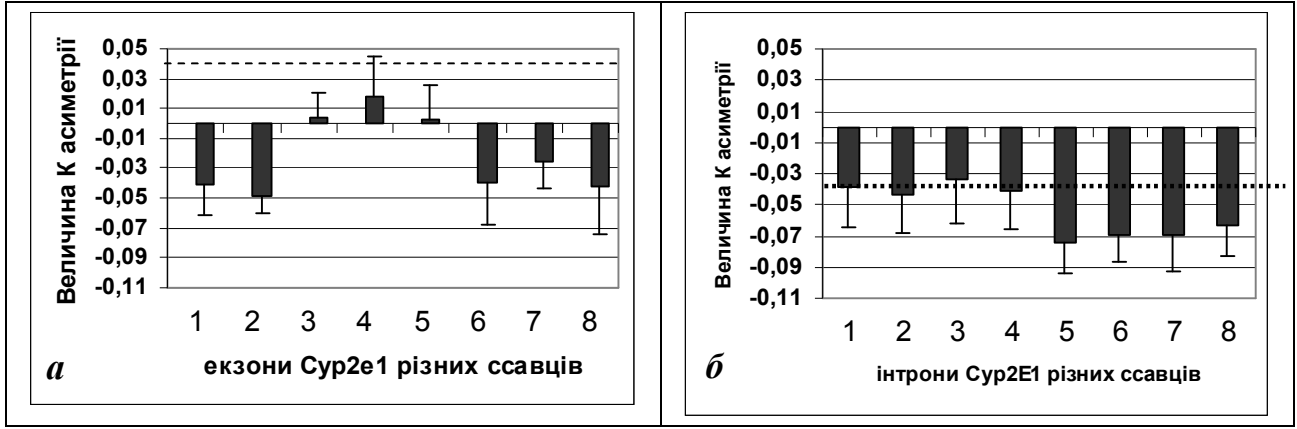

**Рис. 8. Величини К$_{асим}$ в інтронах (*а*) і екзонах (*б*) генів *Cyp2e1* людини та семи ссавців: 1-*Mus musculus*; 2- *Rattus norvegicus*; 3- *Pan troglodytes*; 4-*Homo sapiens*; 5- *Equus caballus*; 6- *Canis lupus*; 7- *Sus scrofa*; 8- *Bos taurus*. Пунктиром відзначено середні величини К$_{асим}$ для 10839 генів людини**

З рис. 8 *а* видно, що в практично всіх екзонів *Cyp2e1* проаналізованих ссавців величина кето-амінної асиметрії є негативною, а у людини – позитивною. А в інтронів ця величина у *Equus caballus* більша ніж у *Homo sapiens*. На рис. 9 показано залежність величини К$_{асим}$ від еволюційного віку біологічного виду. Приблизний час дивергенції (млн. років) біологічних видів приведено за (Carrol, 1992).

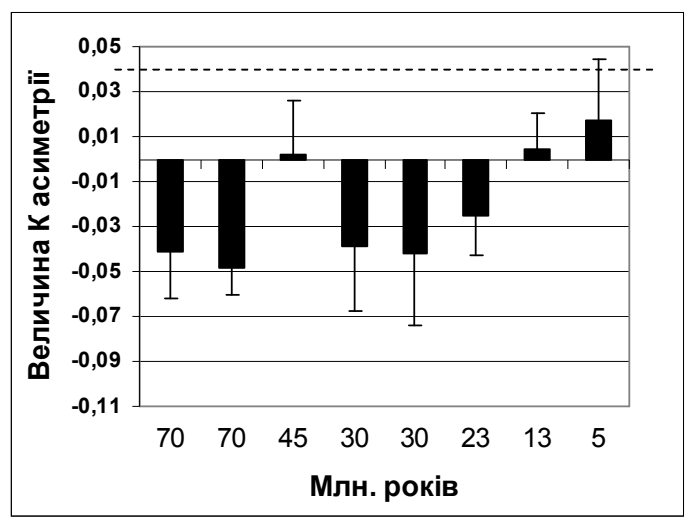

**Рис. 9. Величини К$_{асим}$ в інтронах генів *Cyp2e1* людини та семи ссавців: 70-*Mus musculus*, *Rattus norvegicus*; 45- *Equus caballus*; 30 *Bos taurus*; *Canis lupus*; 23- *Sus scrofa*; 13- *Pan troglodytes* і 5-*Homo sapiens***

Найбільш ймовірною причиною цих співвідношень є накопичення протягом еволюції транзицій CG → TA в інтронах гену *Cyp2e1,* які перебувають під слабкою дією репараційних систем та природного відбору. У результаті проведеного аналізу подібності послідовностей генів цитохрому Р450 2Е1 виявлено, що в інтронах генів *Cyp2e1* найчастішою подією є транзиції GC →AT і CG→TA, які складають 48% від всіх типів однонуклеотидних замін.

Таким чином, показано, що нетрансльовані фрагменти генів мають особливий нуклеотидний склад з переважанням тиміну і гуаніну, в той час як трансльовані ділянки збагачені на цитозин і аденін. У цьому сенсі величина $К_{асим}$ становить інтерес для подальшого дослідження, оскільки може використовуватися як додатковий критерій для оцінки функції нуклеотидної послідовності. Одержані результати є певним етапом дослідження лінійної архітектури геному.

## ВИСНОВКИ

У дисертаційній роботі досліджено композиційний склад геномних послідовностей людини і виявлено залежність нуклеотидних співвідношеннь від структури генів та деякими особливостями їхньої експресії.

1. Розроблено програмний інструментарій оптимізації роботи з нуклеотидними послідовностями з бази NCBI на локальному комп'ютері. Створено базу даних послідовностей генів людини у власному форматі, яка доступна користувачам.

2. Показано суттєву відмінність асиметрій нуклеотидного складу екзонів і інтронів генів людини. В екзонах зафіксовано вищі показники пуриново-піримідинової, а в інтронах – кето-амінної асиметрії.

3. Виявлено залежність величин асиметрій від характеру експресії генів: в інтронах генів, що експресуються конститутивно і виконують базові клітинні функції, спостерігаються найвищі величини асиметрій нуклеотидного складу. Запропоновано застосовувати показник кето-амінної асиметрії як додатковий критерій оцінки функції невідомих послідовностей.

4. Проведено порівняльний аналіз послідовностей генів цитохрому Р450 2Е1 людини та деяких ссавців. На основі подібності смислових ділянок *Cyp2e1* побудовано кладистичне дерево. Виявлено близькість послідовностей *Cyp2e1 Homo sapiens* із послідовностями *Mus musculus* і *Rattus norvegicus*. Це вказує, що проаналізовані гени *Cyp2e1* мають спільну послідовність попередника.

## ПЕРЕЛІК НАУКОВИХ РОБІТ, ОПУБЛІКОВАНИХ ЗА ТЕМОЮ ДИСЕРТАЦІЇ

**АНОТАЦІЯ**

**Дуплій Д. Р. Взаємозв'язок нуклеотидного складу послідовностей генома людини зі структурою та особливостями експресії генів. – Рукопис.**

Дисертація на здобуття наукового ступеня кандидата біологічних наук за спеціальністю 03.00.03 – молекулярна біологія. – Інститут молекулярної біології та генетики НАН України, Київ, 2010.

Дисертацію присвячено дослідженню взаємозв'язку функції елементів геному людини з їхнім нуклеотидним складом. Як характеристику нуклеотидних співвідношень вибрано показники асиметрій нуклеотидного складу. Показано суттєву відмінність асиметрій нуклеотидного складу екзонів і інтронів генів людини. В екзонах виявлено вищі показники пурин-піримідинової, а в інтронах – кето-амінної асиметрії. Знайдено залежність величин асиметрій від характеру експресії генів: найвищі величини асиметрій нуклеотидного складу спостерігаються в інтронах генів, що експресуються конститутивно і виконують базові клітинні функції. Зафіксовано тенденцію накопичення транзицій AT →GC і CG → TA в інтронах, тобто на ділянках менш ефективної дії репараційних систем. Проведено порівняльний аналіз послідовностей генів цитохрому P450 2E1 людини з представниками ссавців. На основі подібності смислових ділянок *Cyp2e1* побудовано кладистичне дерево.

Розроблено програмний інструментарій для аналізу послідовностей бази NCBI і створено базу даних генів людини у власному форматі.

**Ключові слова:** геном людини, асиметрії нуклеотидного складу, експресія генів, цитохром P450 2E1, інтрони, екзони, точкові мутації.

**АННОТАЦИЯ**

**Дуплий Д. Р. Взаимосвязь нуклеотидного состава последовательностей генома человека со структурой и особенностями экспрессии генов. – Рукопись.**

Диссертация на соискание ученой степени кандидата биологических наук по специальности 03.00.03 – молекулярная биология. – Институт молекулярной биологии и генетики НАН Украины, Киев, 2010.

Диссертационная работа посвящена исследованию взаимосвязи функции элементов генома человека с их нуклеотидным составом. В пределах полных геномов или хромосом выполняется правило равенства комплементарных нуклеотидов вдоль одной цепи. В то же время на более коротких участках отношение разности к сумме комплементарных нуклеотидов отличается от единицы, то есть наблюдается асимметрия нуклеотидного состава. В основе этого феномена лежит асимметрия процессов матричного синтеза проходящих в лидирующей и отстающей цепях. Однонуклеотидные замены, возникающие во время транскрипции и репликации попадают под контроль репарационных систем, которые действуют более эффективно в транслируемых областях.

В качестве характеристики нуклеотидных соотношений выбраны показатели асимметрий нуклеотидного состава. Охарактеризованы четыре типа асимметрий нуклеотидного состава: аденин-тиминовая (А-Т)/(А+Т), цитозин-гуаниновая (C-G)/(C+G), пурин-пиримидиновая (A+G-C-T)/(A+G+C+T), и кето-аминная (G+T-C-A)/(A+G+C+T).

Показано существенное различие нуклеотидного состава экзонов и интронов генов человека. В экзонах обнаружены более высокие показатели пурин-пиримидиновой (то есть преобладание количеств A и G над T и C), а в интронах – кето-аминная (то есть преобладание количеств G и T над A и C) асимметрии.

Сформулированы различия асимметрий нуклеотидного состава в экзонах и интронах в виде систем соотношений: $A > T$; $G \geq C$; $A+G > C+T$; $A+C \geq G+T$ для экзонов и $A < T$; $C \leq G$; $A+G \leq C+T$; $A+C < G+T$ для интронов. Найденные различия, демонстрируют обогащение интронных областей G и T.

Для состава интронов характерно неравенство $A+C < G+T$, которое следует из рассчитанной кето-аминной асимметрии. Левая сторона неравенства содержит наиболее часто мутирующие нуклеотиды по данным литературы. Поэтому показатель кето-аминной асимметрии заслуживает отдельного внимания, так как выявляет в интронах преобладание G+T, чего экзонах не наблюдается. Этот факт может быть следствием накопления пуриновых транзиций AT→GC и пиримидиновых CG→TA в интронах, которые находятся под менее сильным действием репарирующих систем и естественного отбора.

Обнаружено, что существенный вклад в величины асимметрий нуклеотидного состава обусловлен характером экспрессии генов. Наибольших величин асимметрии нуклеотидного состава, наблюдаются в интронах конститутивно экспрессирующихся генов, выполняющих базовые клеточные функции, а наименьшие — в прилегающих к генам, нетранскрибируемых областях.

Выявлена зависимость величин асимметрий нуклеотидного состава от характера экспрессии генов по литературным данным анализа ДНК-чипов. Получена положительная корреляция величины кето-аминной асимметрии с уровнем экспрессии конститутивно экспрессирующихся генов, выполняющих базовые клеточные функции в линии сперматоцитов человека. Предложено использовать показатель кето-аминной асимметрии как дополнительный критерий оценки функции неизвестных последовательностей.

Проведен сравнительный анализ последовательностей генов цитохрома P450 2E1 человека с представителями млекопитающих. Построено кладистическое дерево на основе сходства смысловых участков генов *Cyp2e1*. Выявлено, что последовательность гена *Cyp2e1* претерпела несколько дивергенций, в частности, имеются независимые пути развития последовательностей генов грызунов, приматов и прочих млекопитающих. Обнаружена близость последовательности *Cyp2e1 Homo sapiens* с последовательностями *Mus musculus* и *Rattus norvegicus*. Анализ однонуклеотидных подстановок гена *Cyp2e1* человека и семи млекопитающих, показал, что в локусах наибольшего сходства интронов замены встречаются чаще, чем в экзонах. В частности, транзиция AT→GC наблюдается чаще, чем TA→CG, а GC→AT чаще, чем CG→TA. Показана тенденция к накоплению в течение эволюции транзиций CG→TA в интронах генов *Cyp2e1*. Величины кето-аминной асимметрии экзонов *Cyp2e1* человека и семи млекопитающих приблизительно равны, что объясняется высокой степенью гомологии их последовательностей. В отличие от этого величина кето-аминной асимметрии интронов *Cyp2e1* имеет в отрицательное значение у *Mus musculus* и *Bos Taurus,* а у *Pan troglodytes* и *Homo sapiens* положительное. Разработан программный инструментарий для получения и анализа последовательностей базы NCBI, с помощью которого создана собственная база последовательностей экзонов, интронов, а также 3'- и 5'- прилегающих последовательностей генов человека. Разработанный инструментарий может быть применен для анализа других геномов. Полученные результаты могут быть использованы для дальнейших исследований в таких областях как генетика, сравнительная геномика, биоинформатика и некоторые другие.

Результаты проведенного исследования подтверждают гипотезу, касающуюся одного из важных механизмов спонтанного мутагенеза, сопряженного с транскрипцией. Расхождение цепей ДНК и «свободное» их состояние во время процессов матричного синтеза, провоцирует спонтанное дезаминирование цитозина и аденина. В результате исследования обнаружено обогащение тимином и гуанином транскрибируемых, но нетранслируемых областей, что указывает связь AT→GC и CG→TA замен с процессом транскрипции.

**Ключевые слова:** геном человека, асимметрии нуклеотидного состава, экспрессия генов, цитохром Р450 2Е1, интроны, экзоны, точечные мутации.

## SUMMARY

**Relationships among the nucleotide content of human genome sequence, gene structure, and gene expression features. – Manuscript.**

Dissertation to attain the rank of the Candidate of Biological Sciences by speciality 03.00.03 – molecular biology. Institute of Molecular Biology and Genetics of Ukrainian National Academy of Sciences, Kyiv, 2010.

The Dissertation is focused on the studies of associations between functional elements in human genome and their nucleotide structure. The asymmetry in nucleotide content (skew, bias) was chosen as the main feature for nucleotide structure. A significant difference in nucleotide content asymmetry was found for human exons vs. introns. Specifically, exon sequences display bias for purines (i.e., excess of A and G over C and T), while introns exhibit keto-amino skew (i.e. excess of G and T over A and C). The extents of these biases depend upon gene expression patterns. The highest intronic keto-amino skew is found in the introns of housekeeping genes. In the case of introns, whose sequences are under weak repair system, the AT→GC and CG→TA substitutions are preferentially accumulated.

A comparative analysis of gene sequences encoding cytochrome P450 2E1 of *Homo sapiens* and representative mammals was done. Cladistic tree on the basis of coding sequences similarity of the gene *Cyp2e1* was constructed.
A new programming tools of NCBI database sequence mining and analysis was developed, resulting in construction of a own database.
**Keywords:** human genome, nucleotide content asymmetries (skew, bias), gene expression, cytochrome P450 2E1, introns, exons, point mutations.